\documentclass[12pt]{article} 
\usepackage{url,hyperref,lineno,microtype,subcaption,amssymb,amsmath}
\usepackage[onehalfspacing]{setspace}
\usepackage{graphicx}
\usepackage[english]{babel}

\begin{document}

\title{What we talk about\\when we talk about blazars?}
\author{Luigi Foschini\\
\small{INAF -- Osservatorio Astronomico di Brera, Merate (LC), Italy}}
\date{\today}
\maketitle

\begin{abstract}
After the discovery of powerful relativistic jets from Narrow-Line Seyfert 1 Galaxies, and the understanding of their similarity with those of blazars, a problem of terminology was born. The word blazar is today associated to BL Lac Objects and Flat-Spectrum Radio Quasars, which are somehow different from Narrow-Line Seyfert 1 Galaxies. Using the same word for all the three classes of AGN could drive either toward some misunderstanding, or to the oversight of some important characteristics. I review the main characteristics of these sources, and finally I propose a new scheme of classification.
\end{abstract}

\section{Introduction}
The title is borrowed from Haruki Murakami's \emph{What we talk about when we talk about running}, who, in turn, borrowed it from Raymond Carver's \emph{What we talk about when we talk about love}. Far from competing with those two outstanding authors, I just would like to draw the attention on some recent discoveries about relativistic jets, and how to include them into the unified model of active galactic nuclei (AGN) with jets. I would like to underline that this is not a challenge to the unified model, but rather the request of an evolution and an improvement. 

It is not the first time that there is an evolution in the terminology of this type of cosmic sources. This should not be looked as a mere fashion about words. It is true that physical objects exist independently on how we name them, but it is also true that using the most proper words makes it easier to study them, by avoiding to remain stuck on a swamp of fake problems and misleading questions. When Gregorio Ricci Curbastro and Tullio Levi-Civita proposed the tensor calculus, many other mathematicians rejected it, because they thought it was just a mere rehash of old maths. When speaking about Ricci Curbastro, Luigi Bianchi told that he preferred to find new things with old methods, rather than to find old things with new methods (cited in \cite{TOSCANO}). On the opposite, Henri Poincar\'e wrote that a proper notation in mathematics has the same importance of a good classification in natural science, because it allows us to connect each other many events without any apparent link (cited in \cite{BOTTAZZINI}).

Back to the topic of this essay, I would like to remind some past changes in terminology about relativistic jets. In 1978, Ed Spiegel proposed the term \emph{blazar} as a contraction of the words BL Lac Objects and Optically Violently Variable Quasars (OVV) \cite{ANGELSTOCKMAN}; in 1994-1995, Paolo Padovani and Paolo Giommi proposed to rename radio-selected BL Lac objects (RBL) as low-energy cutoff BL Lacs (LBL), and X-ray selected BL Lac objects (XBL) changed to high-energy cutoff BL Lacs (HBL) \cite{GP94,PG95}; also the Fanaroff-Riley classes of radio galaxies changed to low- and high-excitation radio galaxies (LErG, HErG) \cite{HL79,LAING94,BUTTIGLIONE}. In his opening talk at the conference \emph{Quasar at all cosmic epochs} (Padova, April 2-7, 2017), Paolo Padovani proposed to stop using radio loud/quiet terms and to start speaking about jetted AGN or not. I was less severe in my thoughts on radio loudness some years ago \cite{FOSCHINI7}, although I agreed with Padovani. It is time to be resolute in changing terminology. Also Martin Gaskell wrote: ``I tell students that classification is one of the first step in science. As science progresses, however, I believe that we need to move toward physically meaningful classification schemes as soon as possible. To achieve this, we need to be willing to modify our definitions, or else we can impede progress'' (cited in \cite{QUASAR50}). This means to move from a purely observational classification to a terminology with more physical grounds. It should be needless to say, but before establishing the type of a cosmic source, it is necessary to study it. A simple measure is the easy way, but it is also the most prone to errors. 

Today, AGN with jets are unified according to the scheme by Urry \& Padovani \cite[Table 1]{UP95}, which in turn summarises many years of contributions from pioneers (see the historical review in \cite{QUASAR50}). Urry \& Padovani's scheme has its pillars in three main factors: viewing angle, optical spectrum, radio emission. They also suggested a fourth factor, the black hole spin, which should be greater for jetted AGN. 

With reference to jetted AGN only, the Urry \& Padovani's unified model can be divided into two main classes and four subclasses on the basis of viewing angle and optical spectrum \cite{UP95}: 

\begin{enumerate}
\item blazar (small viewing angle, beamed sources):
\begin{enumerate}
\item flat-spectrum radio quasar (FSRQ), prominent emission lines in the optical spectrum;
\item BL Lac Object, weak emission lines or featureless continuum in the optical spectrum;
\end{enumerate}
\item radio galaxies (large viewing angles, unbeamed sources):
\begin{enumerate}
\item HErG, prominent emission lines in the optical spectrum;
\item LErG, weak emission lines or featureless continuum in the optical spectrum;
\end{enumerate}
\end{enumerate}

The mass of the central spacetime singularity was generally in the range $\sim 10^{8-10}M_{\odot}$ \cite{BUTTIGLIONE,GG10,TADHUNTER16}, which seemed to fit well with the elliptical galaxies hosting this type of cosmic sources \cite{OI16}. The limited range meant to neglect the mass when scaling the jet power. Therefore, the main factor regulating the electromagnetic emission became the electron cooling, which is the basis of the so-called \emph{blazar sequence} \cite{FOSSATI,GG98}. The observed blazar sequence indicated that the spectral energy distribution (SED) of high-power blazar (FSRQs) had the synchrotron and the inverse-Compton peaks at infrared and MeV-GeV energies, respectively, while that of low-power sources (BL Lac Objects) had the peaks shifted to greater energies (UV/X-rays and TeV, respectively) \cite{FOSSATI}. This was explained as different cooling of relativistic electrons due to different environment, rich of photons or not (physical blazar sequence, \cite{GG98}). In addition, since no other jetted AGN with smaller masses were known, it was thought that the generation of a relativistic jet required a minimum black hole mass \cite{LAOR,CHIABERGE}. 

Truly speaking, the lack of small-mass jetted AGN was a selection bias. For example, in 1979, Miley and Miller \cite{MILEY} studied a sample of 34 quasars with $z<0.7$: their sample included also objects with small black hole mass, which resulted to have compact radio morphology. In 1986, Wills and Browne \cite{WILLS} studied a sample of 79 quasars with the same redshift range, but selecting only bright sources (mag~$<17$): small-mass objects disappeared. Therefore, jets from small-mass AGN were known at least since seventies, but they were disregarded, likely because of the poor instruments sensitivity. The recent technological improvements resulted in an increase of the cases of powerful jets hosted in spiral galaxies (hence, small mass of the central black hole) \cite{KEEL,MORGANTI}. Also recent surveys showed that disk/spiral hosts are not just an exception, but they could be a significant fraction of jetted AGN \cite{INSKIP,COZIOL}. Particularly, Coziol et al. \cite{COZIOL} confirmed the results of \cite{MILEY}: small-mass compact objects have generally weak, and compact radio jets. 

\section{High-Energy Gamma Rays from Narrow-Line Seyfert 1 Galaxies}
The turning point occurred in 2009, with the detection of high-energy $\gamma$ rays from Narrow-Line Seyfert 1 Galaxies (NLS1), thus providing evidence of powerful relativistic jets from small-mass AGN \cite{LAT1,LAT2,LAT3,FOSCHINI8} (see also \cite{FOSCHINI1} for a historical review). NLS1s do have small-mass central black holes ($\lesssim 10^{8}M_{\odot}$), high accretion luminosity (close to the Eddington limit), prominent optical emission lines, but relatively weak jet power, comparable to BL Lac Objects \cite{FOSCHINI2}. Kinematics of radio components revealed superluminal motion ($\sim 10c$ \cite{LISTER}; see also \cite{ANGELAKIS,LAHTEENMAKI} for more information about radio properties), while infrared colors indicated an enhanced star formation activity \cite{CACCIANIGA}. The host galaxy is not yet clearly defined: there is evidence that NLS1 without jets are hosted by spiral galaxies, but $\gamma-$ray NLS1 are still poorly known. However, early observations of a handful of sources point to disk galaxy hosts\footnote{Two opposite interpretation were proposed for FBQS~J1644$+$2619, a spiral barred host \cite{OI17}, and an elliptical galaxy \cite{DAMMANDO}. However, the former observation seems to be more reliable, because done with a much better seeing than the latter.}, the result of either a recent merger or a secular accretion \cite{ZHOU,ANTON,HAMILTON,LEONTAVARES,KOTILAINEN,OI17}. 

All the observed characteristics of NLS1s suggested that this class of AGN could be the low-mass tail of the quasars distribution \cite{LAT1,FOSCHINI2}. Indeed, as proved by Berton et al. \cite{BERTON1}, the NLS1s luminosity function matches that of FSRQs. The parent population of jetted NLS1s could be that of Compact Steep Spectrum (CSS) HErG \cite{BERTON1,BERTON2}. This fits well with the idea that NLS1s are quasars at the early stage of their evolution or rejuvenated by a recent merger \cite{MATHUR}. 

However, I would like to underline that it is not a matter of NLS1s only, but of small-mass AGN. Recent surveys with \emph{Fermi/LAT} \cite{SHAW,FOSCHINI3} and the \emph{Sloan Digital Sky Survey} \cite{BEST} indicated that jetted AGN with small-mass black holes are not restricted to NLS1s-type AGN. The exact observational classification is not the point, but what is important is the relatively small mass of the central spacetime singularity. This confirms once again that the mass threshold to generate the relativistic jet in AGN was just an observational bias. 

\begin{figure}[h!]
\begin{center}
\includegraphics[width=\linewidth]{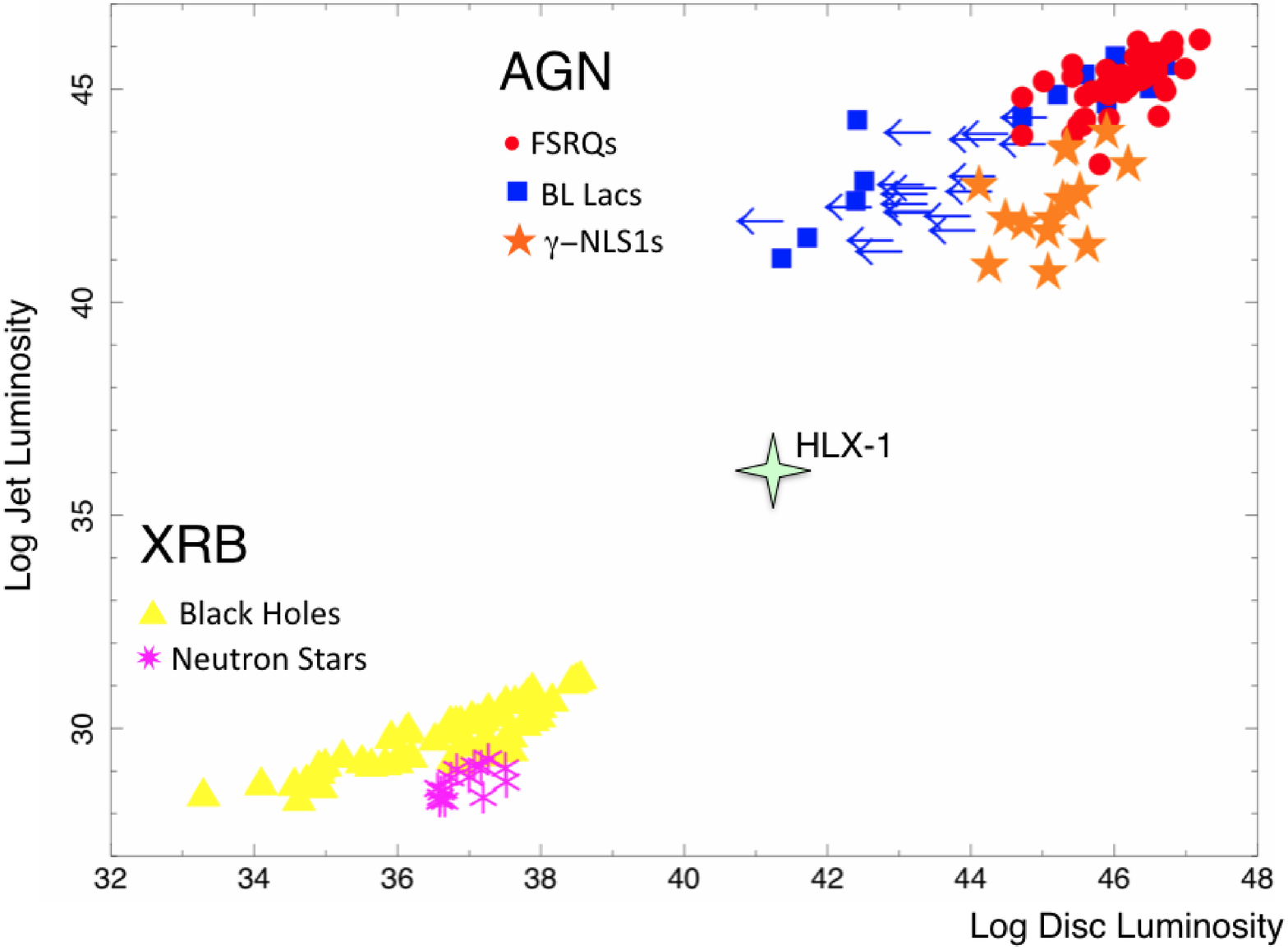}
\end{center}
\caption{Unification of relativistic jets. The AGN sample includes FSRQs (red circles), BL Lac Objects (blue squares or arrows for upper limits in disk luminosity), and NLS1s (orange stars). The XRB sample includes different states of some stellar-mass black holes (yellow triangles) and accreting neutron stars (magenta asterisks). HLX-1 is the only known jetted ultraluminous X-ray source with a reliable intermediate-mass black hole. Adapted from \cite{FOSCHINI5}.}\label{fig:1}
\end{figure}

\section{The unification of relativistic jets}
Removing the mass-threshold bias has the important consequence of the unification of relativistic jets from AGN and from Galactic X-ray binaries (XRBs) \cite{FOSCHINI4,FOSCHINI5}. As shown in Fig.~\ref{fig:1}, in a jet power vs disk luminosity graph, NLS1s populate the previously missing branch of small-mass highly-accreting compact objects, the analogous of accreting neutron stars for the XRBs sample (see also \cite{PALIYA} for a larger sample of AGN). By applying the scaling relationships elaborated by Heinz \& Sunyaev \cite{HEINZ}, it is possible to merge the AGN and XRBs populations. A residual dispersion of about three orders of magnitudes remain (see \cite{FOSCHINI4,FOSCHINI5}): measurements errors could account for about one order of magnitude, while the remaining two could likely to be due to the spin of the compact object \cite{HEINZ,SIBHS}, whose measure is still missing or largely unreliable (see also \cite{FOSCHINI6}).

\section{Implications of the unification}
\begin{figure}[t!]
\begin{center}
\includegraphics[width=\linewidth]{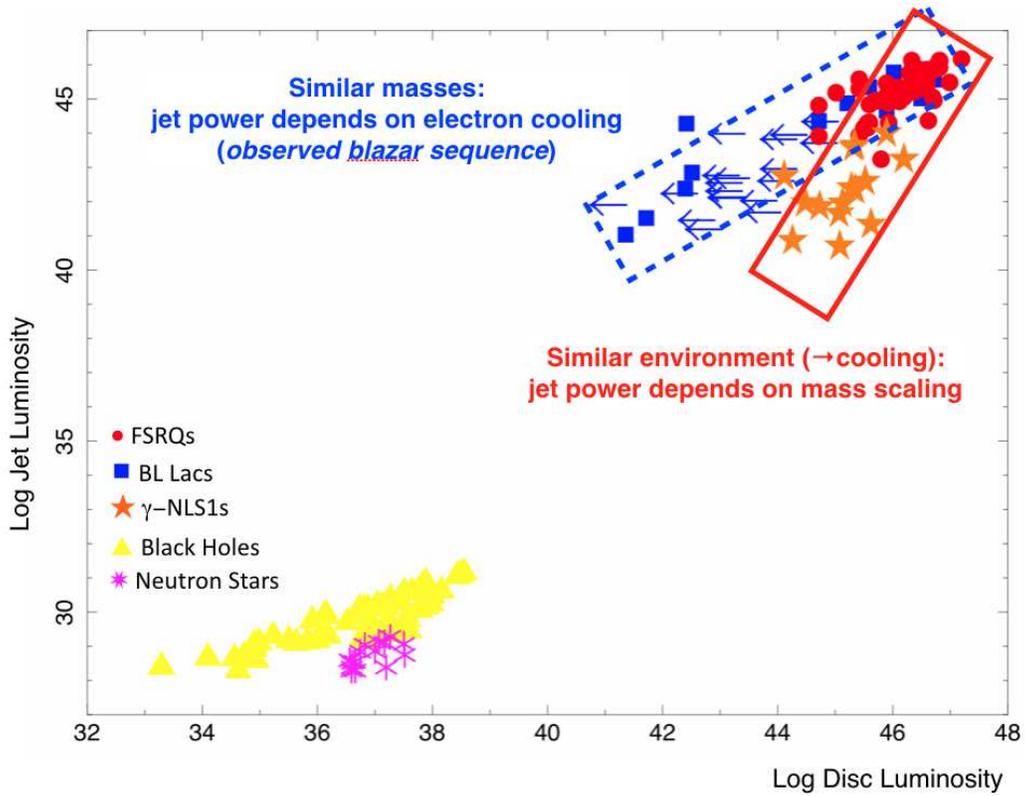}
\caption{Implications of the unification of relativistic jets: different branches, different mechanisms.}\label{fig:2a}
\end{center}
\end{figure}

\begin{figure}[t!]
\begin{center}
\includegraphics[width=\linewidth]{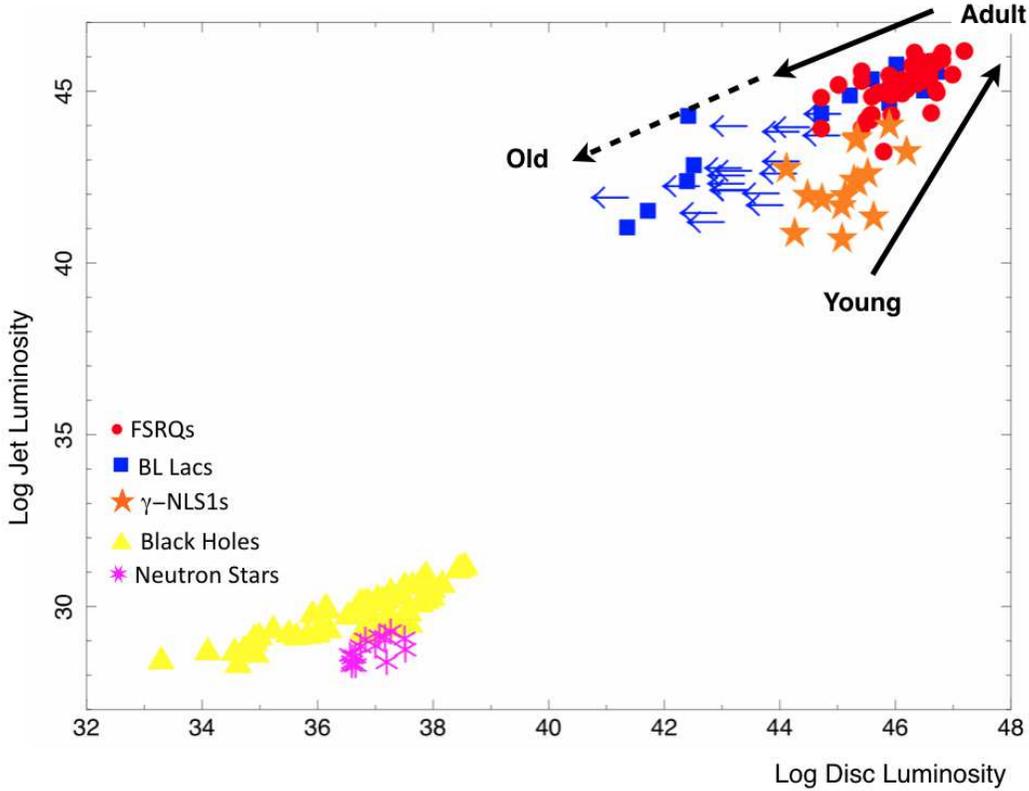}
\caption{Implications of the unification of relativistic jets: cosmological evolution.}\label{fig:2b}
\end{center}
\end{figure}

Having proved the Heinz \& Sunyaev's scaling theory \cite{HEINZ}, the jet power vs disk luminosity graph could be used also to understand and to visualise some implications of the unification of relativistic jets (Fig.~\ref{fig:2a},\ref{fig:2b}). Each population has two branches, depending on the main factor driving the changes in the jet power. The dashed blue rectangle in Fig.~\ref{fig:2a} summarises the blazar sequence \cite{FOSSATI,GG98}: the black hole masses of blazars are within about one, maximum two, orders of magnitudes, and, therefore, the main changes in the jet power are driven by the electron cooling in different environments. The red rectangle in Fig.~\ref{fig:2a} refers to similar environments (FSRQs and NLS1s), but largely different masses, which in turn implies that the main change in the jet power is due to the mass of the central black hole \cite{HEINZ}. The small-mass black hole is necessary to explain the weak jet power of NLS1s, which is comparable with BL Lac Objects \cite{FOSCHINI2}: as the environment of NLS1s is rich of photons like FSRQs, a large black hole mass would mean that relativistic electrons of the jet do no cool efficiently with so many photons, thus contradicting a basic physical law. Indeed, BL Lac Objects have weak jet and large masses, but their environment is photon-starving (see also \cite{FOSCHINI6}). The NLS1s branch (red rectangle) also prove that the observational blazar sequence (the dashed blue rectangle only \cite{FOSSATI}) was due to a bright-source selection bias, although the physical blazar sequence \cite{GG98} remains valid, as it simply refers to how relativistic electrons cool depending on photon availability. 

When comparing the AGN with XRBs samples, one can note that the blazar sequence corresponds somehow to the stellar-mass black hole states, but on different time scales. Galactic black holes evolve on human time scales: it is possible to observe a state change of a black hole along months/years of observations. A quasar requires some billion of years to swallow most of the available interstellar gas and to become a BL Lac Object \cite{CAVALIERE,BOTTCHER,MARASCHI}. This opens another implication, namely the cosmological evolution (Fig.~\ref{fig:2b}). The blazar evolutionary sequence \cite{CAVALIERE,BOTTCHER,MARASCHI} suggested that quasar are young AGN, which become BL Lac Objects as they grow. This scenario has to be updated now by adding also NLS1s, which are thought to be a low-redshift analogous of the early quasars \cite{MATHUR,BERTON2}. Therefore, one could think at the sequence NLS1s~$\rightarrow$~FSRQs~$\rightarrow$~BL Lac Objects, going from small-mass highly-accreting to large-mass poorly-accreting black holes, as different stages of the cosmological evolution of the same type of source (young~$\rightarrow$~adult~$\rightarrow$~old, Fig.~\ref{fig:2b}). This view implies that BL Lac Objects have the largest masses, being (perhaps) the final stage of the cosmic lifetime, at odds with the results of some surveys \cite{GG10}. Again, if one removes the bright sources selection bias, it is possible to find that indeed BL Lacs/LErGs do have masses larger than FSRQs/HErGs \cite{BEST}.

On the other side of the evolution, it is worth noting the presence of strong star formation in NLS1s, with infrared properties similar to UltraLuminous InfraRed Galaxies (ULIRGs) \cite{CACCIANIGA}. This points also to some link to the very birth of a quasar and its jet, which in turn could be an essential angular momentum relief valve to enhance the accretion \cite{JOLLEY}. ULIRGs as early quasar stage were already studied by Sanders et al. \cite{SANDERS1,SANDERS2} and it is interesting to note the presence in his sample of both the NLS1 and the quasar prototypes (I Zw 1, and 3C 273, respectively). 
 
It is also worth noting the application of the same sequence to the XRBs population, which implies a transition from accreting neutron stars to stellar-mass black holes \cite{NSBH1,NSBH2,NSBH3}. 

\section{Conclusion: a renewed unified scheme}
\begin{figure}[h!]
\begin{minipage}[b]{.5\linewidth}
\centering\includegraphics[width=6.5cm]{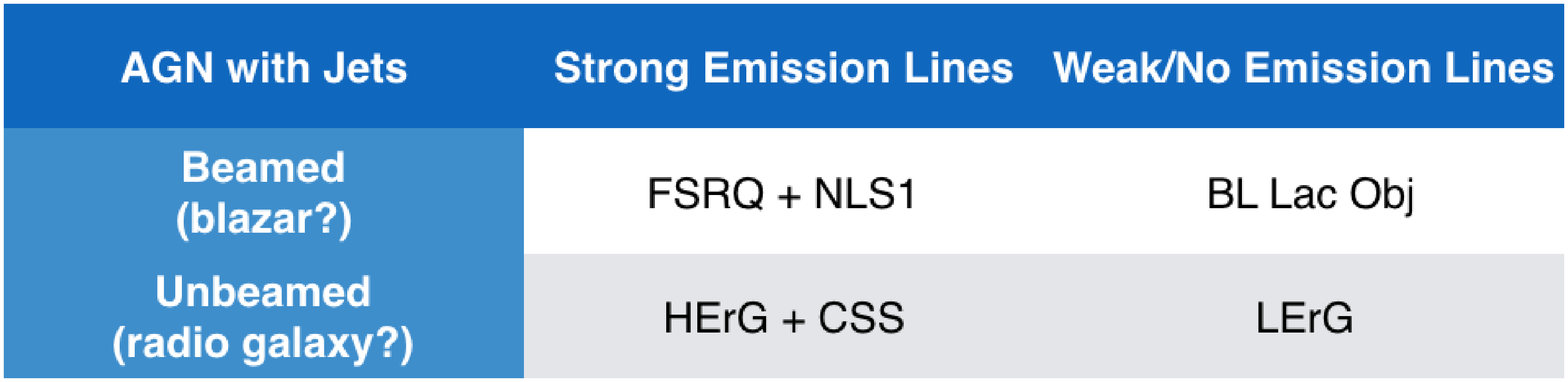}
\subcaption{Urry \& Padovani's scheme updated.}\label{fig:3a}
\end{minipage}%
\begin{minipage}[b]{.5\linewidth}
\centering\includegraphics[width=6.5cm]{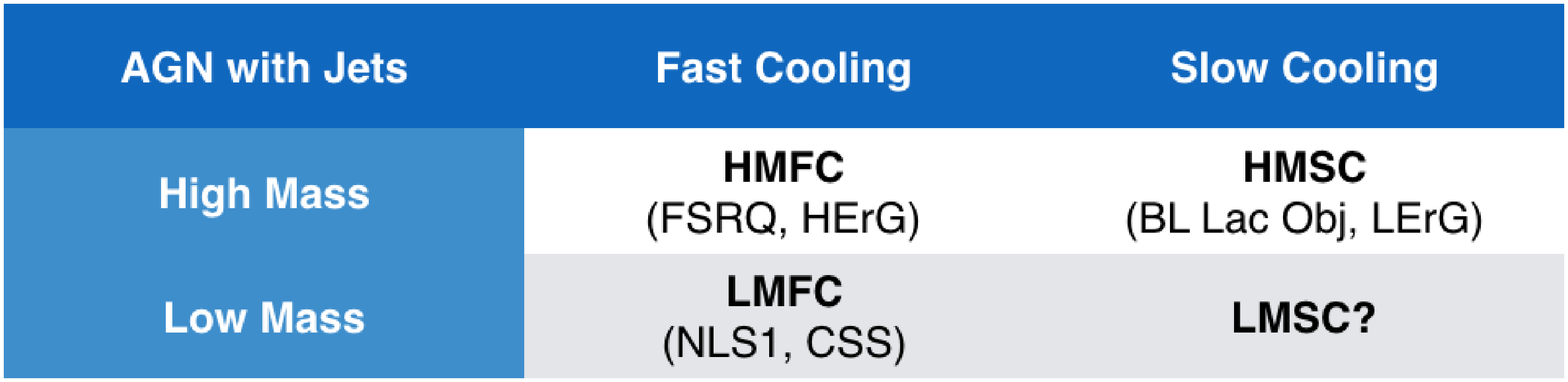}
\subcaption{Physical unification scheme.}\label{fig:3b}
\end{minipage}
\caption{Changing terms in the unification scheme.}\label{fig:3}
\end{figure}

The Urry \& Padovani's scheme \cite{UP95} updated with the addition of NLS1s and their parent population of CSS/HErG is shown in Fig.~\ref{fig:3a}. However, this generates some problem in terminology. The words blazar and radio galaxy indicate a certain type of cosmic source characterised by a high black hole mass and hosted by an elliptical galaxy. The easy addition of NLS1s and CSS to the above scheme risks to hide important information, as outlined in the previous section (different black hole mass, different host, ...). This is not a negligible detail: remind the misleading research directions caused by the bright sources selection bias, such as the threshold in the jet generation and the observed blazar sequence. Martin Gaskell wrote: ``When you attach different classification to things, it is all too easy to get convinced that they are different things.'' (cited in \cite{QUASAR50}). On the opposite, if you attach the same name to different things, it is all too easy to get convinced that they are the same thing. Therefore, on one side, we need to unify jetted AGN, but, on the other side, we need to keep some information about the roots of this unification to understand the physical processes driving the observational characteristics. The jets of AGN and XRBs are similar, but their power depends on the mass of the compact object, its spin, and its accretion (environment). It is important to note that Fig.~\ref{fig:1} was built by using the jet power corrected for beaming. Indeed, it places on the same plane both beamed AGN and XRBs, which are not so beamed, as it is quite difficult for a Galactic jet to point toward the Earth, being both on the same equatorial Galactic plane\footnote{Galactic compact objects with jets are named microquasars, but there is no such thing as a microblazar, i.e. a Galactic jet pointed toward the Earth.}. The addition of HErG/LErG/CSS sources would not change the two-branches structure for each population. Therefore, it should be possible to drop also the distinction beamed/unbeamed. From a physical point of view, the two most important factors in scaling the jet power are the mass of the compact object and the nearby environment (for the electron cooling), which in turn depends on the accretion. As already stated, the spin determines a larger dispersion only \cite{HEINZ,SIBHS}. Therefore, a more physical-based unification could be set up by dividing the sources depending on the mass and on the cooling only (Fig.~\ref{fig:3b}). The dividing mass is $\sim 10^8M_{\odot}$ not because of historical reasons, but because no BL Lac Object with small mass is known. Indeed, I have left a question mark on the LMSC (Low Mass Slow Cooling) cell. Current BL Lacs should be the latest stage of the cosmological evolution of jetted AGN, and, therefore, a small-mass BL Lac would mean that there was no evolution. Did such AGN have no matter enough for accretion? As there are other small-mass AGN, which are not necessarily NLS1-type \cite{SHAW,FOSCHINI3,BEST}, it would be interesting to understand if some of them have a photon-starving environment. Perhaps, one intriguing case could be PKS~2004$-$447 ($z=0.24$) that showed observational characteristics somehow different from the other jetted NLS1s \cite{LAT3,KREIKENBOHM,SCHULZ}. There was also some disagreement on its classification as NLS1s, on the basis of the weakness of the FeII multiplets \cite{GALLO,KOMOSSA}. It is interesting to point out that it is the only NLS1 (orange star) in the region of BL Lac Objects (blue squares or arrows) in Fig.~\ref{fig:1}.  

The same terminology adapts well also to the XRB population: in this case, the dividing mass should be $\sim 3M_{\odot}$, which is the minimum for a Galactic black hole. Also in this case, the LMSC cell remains with a question mark, but the question is more intriguing, because of shorter time scales. Could it be filled by pulsars? Similar questions on AGN evolution apply here.

I stop here, because these are questions on which I have not thought enough yet. But, I think it is important to stress the different view offered by a terminology change built on more physical ground, rather than to focus on observational details.

\section*{Funding}
This work has been partially supported by PRIN INAF 2014 ``Jet and astro-particle physics of gamma-ray blazars'' (PI F. Tavecchio).

\bibliographystyle{ieeetr}
\bibliography{foschini}
\end{document}